\documentclass[prc,twocolumn,showpacs,preprintnumbers,amsmath,amssymb]{revtex4}
\usepackage{epsfig}
\usepackage{graphicx}% Include figure files
\usepackage{bm}% bold math
\usepackage{colordvi}
\def\lsim{\buildrel < \over {_{\sim}}}

\newcommand{\beq}{\begin{equation}}
\newcommand{\eeq}{\end{equation}}
\newcommand{\be}{\begin{eqnarray}}
\newcommand{\ee}{\end{eqnarray}}

\begin{document}
%\preprint{APS/123-QED}
%%%%%%%%%%%%%%%%%%%%%%%%%%%%%%%%%%%%%%%%%%%%%%%%%%%%%%%%%%%%%%%%%%%%%%%%%
\title{Effective interaction approach to the Fermi hard-sphere system}
%%%%%%%%%%%%%%%%%%%%%%%%%%%%%%%%%%%%%%%%%%%%%%%%%%%%%%%%%%%%%%%%%%%%%%%%%
\author{Angela Mecca$^{1,2}$}
\author{Alessandro Lovato$^{3,4}$}
\author{Omar Benhar$^{1,2}$}
\author{Artur Polls$^{5}$}
\affiliation
{
$^1$ INFN, Sezione di Roma. I-00185 Roma, Italy\\
$^2$ Dipartimento di Fisica, ``Sapienza'' Universit\`a di Roma. I-00185 Roma, Italy \\
$^3$ Argonne Leadership Computing Facility, Argonne National Laboratory, Argonne, IL 60439, USA\\
$^4$ Physics Division, Argonne National Laboratory, Argonne, IL-60439, USA \\
$^5$ Departament d'Estructura i Constituents de la Mat\`eria. E-08028 Barcelona, Spain \\
}
%%%%%%%%%%%%%%%%%%%%%%%%%%%%%%%%%%%%%%%%%%%%%%%%%%%%%%%%%%%%%%%%%%%%%%%%%
\date{\today}
%%%%%%%%%%%%%%%%%%%%%%%%%%%%%%%%%%%%%%%%%%%%%%%%%%%%%%%%%%%%%%%%%%%%%%%%%
\begin{abstract}
The formalism based on correlated basis functions and the cluster expansion technique has been recently employed to derive an effective interaction 
from a realistic nuclear hamiltonian.  To gauge the reliability of this scheme, we perform a systematic comparison between the results of  its application 
to the Fermi hard-sphere system and the predictions obtained from low-density expansions, as well as from other many-body 
techniques. The analysis of a variety of properties, including the ground state energy, the effective mass and the momentum distribution,  shows that 
the effective interaction approach is remarkably accurate, thus suggesting that it may be employed to achieve a consistent description of the structure 
and dynamics of nuclear matter in the density region relevant to astrophysical applications.
 \end{abstract}
\maketitle
%%%%%%%%%%%%%%%%%%%%%%%%%%%%%%%%%%%%%%%%%%%%%%%%%%%%%%%%%%%%%%%%%%%%%%%%%

%%%%%%%%%%%%%%%%%%%%%%%%%%%%%%%%%%%%%%%%%%%%%%%%%%%%%%%%%%%%%%%%%%%%%%%%%
\section {Introduction}
\label{intro}
%%%%%%%%%%%%%%%%%%%%%%%%%%%%%%%%%%%%%%%%%%%%%%%%%%%%%%%%%%%%%%%%%%%%%%%%%

Approaches based on effective interactions are widely used to study the properties of 
strongly interacting many-body systems, when the bare interaction between the constituents cannot be treated 
in perturbation theory using the basis states describing non interacting particles \cite{EI1,EI2}. 

%Theoretical studies of nuclear matter have shown that 
Effective interactions specifically designed to reproduce the bulk properties of nuclear matter (see, e.g. Refs.~\cite{skyrme1,skyrme2}), while 
being remarkably successful in a number of instances, 
fail to provide a quantitative account of nucleon-nucleon (NN) scattering in the nuclear medium, 
the understanding of which is needed for the description of non-equilibrium properties \cite{Gof4,lombardo}. 
The results reported in Ref. \cite{Gof4} clearly show that 
the determination of the shear viscosity and thermal conductivity of pure neutron matter, relevant to  many
astrophysical applications \cite{ander,reddy}, requires effective 
interactions derived from {\em ab initio} microscopic approaches, capable of explaining 
the observed NN scattering data in the zero-density limit \cite{Gof4}.

The authors of Refs. \cite{shannon,BV} have developed a procedure to determine the effective interaction   in nuclear matter
using the Correlated Basis Function (CBF) formalism and the cluster expansion technique. 
While this scheme has been thoroughly  tested through comparison between its results and those 
obtained from G-matrix perturbation theory in pure neutron matter \cite{Gof4}, the analysis of a somewhat simpler 
many-body system, several properties of which can be accurately calculated and expressed in analytic form, may provide further insight
in the validity and robustness of the underlying assumptions. 

The nucleon-nucleon (NN) interaction is known to be strongly repulsive at short distances, as clearly indicated by the saturation 
of the charge-density distributions measured by elastic electron-nucleus scattering \cite{densities}. As a consequence, the Fermi hard-sphere fluid, i.e. a system of point-like spin one-half particles interacting through the
potential
\begin{equation}
\label{potential}
v(r) = \left\{
\begin{array}{cc}
\infty & r < a \\
0       & r > a 
\end{array}
\right. \ ,
\end{equation}
has been long recognised as a valuable model for investigating concepts and approximations employed to study the properties of 
nuclear matter.

In this paper we discuss the derivation of the effective interaction of the Fermi hard-sphere system within the approach of
Ref. \cite{BV}, as well as its application to the calculation of a variety of properties, including the energy per particle, 
the self-energy, the effective mass and the momentum distribution.

As an introduction, Sec. \ref{pert} is devoted to a summary of the results of 
%perturbation theory and
 low-density expansions in powers of the dimensionless parameter $c=k_Fa$, where $a$ is the hard core radius [see Eq. \eqref{potential}] and $k_F$ is the 
 Fermi momentum. In Sec. \ref{veff}, we outline the basics of both CBF theory and the cluster expansion 
technique, needed to obtain the ground state energy and determine the effective interaction.
The perturbative calculation of the self-energy at second order in the 
effective interaction is described Sec.~\ref{selfenergy}, while in Sec. \ref{results} the  
single particle properties resulting from our calculations are reported and compared to those obtained using different approaches.
%, which are compared to those obtained from different approaches. 
Finally, in Sec. \ref{concl} we summarize our findings and state the conclusions.

%%%%%%%%%%%%%%%%%%%%%%%%%%%%%%%%%%%%%%%%%%%%%%%%%%%%%%%%%%%%%%%%%%%%%%%%%
\section{Low-density expansions}
\label{pert}

The expansion of the ground state energy of the quantum mechanical hard-sphere system 
in powers of the dimensionless parameter $c$ was first discussed by Huang and Yang \cite{HY}, who 
were able to derive its terms up to order $c^2$, in the 1950s. 

More recently, Bishop carried out a systematic analysis, including a comparison between results 
obtained using different computational schemes \cite{bishop}. 

The calculation of the ground-state energy exploits the formalism developed to describe 
a scattering process involving two particles interacting through a strongly repulsive potential.
The main element of this approach is the replacement of the bare interaction with the $t$-matrix, 
which amounts to including the contribution of the infinite series of ladder diagrams.
This technique, which in general allows to achieve a fast convergence of perturbative calculations, 
becomes essential when dealing with the hard-core interaction of Eq. \eqref{potential}.

The author of Ref. \cite{bishop} considered two different treatments of two-body scattering 
in a degenerate medium, based on use of  time-ordered Goldstone diagrams or Feynman 
diagrams, yielding the same expression of the ground state energy.

For a hard-sphere system of degeneracy $\nu =4$, the result, obtained including the first four terms 
in the expansion, reads
\begin{align}
\label{pert_en}
\nonumber
E_0  = \frac{k_F^2}{2m} \left[ \frac{3}{5} \right. & \left. + \frac{2}{\pi}c + \frac{12}{35 \pi^2} \left( 11 - 2 \ln 2 \right) c^2  + 0.78c^3 \right. \\
& \left. + \frac{32}{9\pi^3} \left( 4 \pi - 3 \sqrt{3} \right) c^4 \ln c + O(c^4) \right] \ , 
\end{align}
where the linear term describes the effects of forward scattering, the quadratic term takes into account Pauli's exclusion principle, 
and the higher order terms arise from the occurrence of processes involving at least three particles.

The low-density expansion for the single particle spectrum, $e(k)$, and the effective mass, defined as
\begin{align}
\label{def:mstar}
m^\star(k) =  \left( \frac{1}{k} \frac{ d e}{d k} \right)^{-1} \ , 
\end{align}
 are discussed in Ref. \cite{EI1}. The result at $k=k_F$, derived taking into account terms quadratic in $c$ is
\begin{align}
\label{pert_mstar}
\frac{m^\star(k_F)}{m} = 1 + \frac{ 24 }{ 15 \pi^2 }( 7 \ln 2 - 1 ) c^2 \ .
\end{align}
Note that the above equation implies that: i) there are no linear contributions and ii) $[m(k_F)^\star/m] >1$ for all values of $c$.

Perturbative results at order $c^2$ have also been obtained for the momentum distribution, defined as
\begin{align}
\label{def:nk}
n(k) = \langle 0 | a^\dagger_{\bf k} a _{\bf k} | 0 \rangle \ , 
\end{align}
carrying out an expansion in powers of the $t$-matrix in free space \cite{galitskii,gottfried,mahaux}. In the above equation,    
$| 0 \rangle$ denotes the system ground state, while $a^\dagger_{\bf k}$ and  $a _{\bf k}$ are creation and annihilation 
operators, respectively. 

%%%%%%%%%%%%%%%%%%%%%%%%%%%%%%%%%%%%%%%%%%%%%%%%%%%%%%%%%%%%%%%%%%%%%%%%%
\section{The CBF effective interaction}
\label{veff}

Within the CBF  approach, the {\em correlated} states of the hard-sphere system 
are obtained from the non interacting Fermi gas (FG) states through the transformation
\begin{equation}
|n\rangle = \frac{F |n_{FG}\rangle}{ \langle n_{FG}| F^\dagger F |n _{FG} \rangle^{1/2}} \ ,
\end{equation}
where the operator $F$, embodying the correlation structure induced by the 
interaction potential, is written in the form
\begin{equation}
F=\prod_{j>i} f(r_{ij}) \  ,
\end{equation}
with 
\begin{equation}
f(r_{ij} \leq a) = 0 \ \ \ , \ \ \ \lim_{r_{ij} \to \infty} f(r_{ij}) = 1 \ ,
\end{equation}
$r_{ij} = | {\bf r}_i - {\bf r}_j|$ being the interparticle distance. 

In principle, the shape of the two-body correlation function, $f(r)$, at $r>a$ can be determined from functional minimisation of the 
expectation value of the hamiltonian
\begin{equation}
H = T + V = \sum_i \frac{k_i^2}{2m} + \sum_{j>i} v(r_{ij})   \ , 
\label{hamiltonian}
\end{equation}
in the correlated ground state. In the above equation, $k_i = |{\bf k}_i|$, $m$ denotes the particle mass and $v$ is the potential of Eq. \eqref{potential}. 

The effective interaction 
\beq
\label{def:veff2}
V{_{\rm eff}} = \sum_{j>i} v_{\rm eff}(r_{ij}) \ ,
\eeq
is defined by the relation
\beq
\label{def:veff}
\langle H \rangle = \frac{1}{N} \frac{\langle 0 | H | 0 \rangle}{\langle 0 | 0 \rangle} = 
%\langle 0_{FG} | T + V_{{\rm eff}}| 0_{FG} \rangle  \ .
K_{FG} \ + \ \langle 0_{FG} | V_{{\rm eff}}| 0_{FG} \rangle  \ ,
\eeq
where $N$ is the number of particles, and $K_{FG} = 3 k_F^2/10m$ is the expectation value of the kinetic energy in the FG ground state.
Note that the above equation implies that the CBF effective interaction is defined {\em not} in operator form,  but in terms of 
its  expectation value in the Fermi gas ground state. 

The calculations discussed in the following Sections are largely 
based on the assumption---that will be tested comparing our results to those obtained from different many-body approaches---that perturbative calculations
involving matrix elements of $V_{\rm eff}$ between Fermi gas states provide accurate estimates of {\em all} properties of the Fermi hard-sphere system.
%ibutions can be summed up to all orders solving the so-called Fermi Hyper-Netted Chain (FHNC) integral 
%equations \cite{FHNC1,FHNC2}.

%%%%%%%%%%%%%%%%%%%%%%%%%%%%%%%%%%%%%%%%%%%%%%%%%%%%%%%%%%%%%%%%%%%%%%%%%%%%%
\subsection{Cluster expansion formalism}
\label{formalism}
%%%%%%%%%%%%%%%%%%%%%%%%%%%%%%%%%%%%%%%%%%%%%%%%%%%%%%%%%%%%%%%%%%%%%%%%%%%%%
The calculation of matrix elements of any many-body operator between correlated states involves largely irreducible $3N$-dimensional integrations. This problem, that becomes 
quickly intractable for large $N$, can be circumvented expanding the matrix element in a series, the terms of which represent the contributions of subsystems (clusters) 
involving an increasing number of particles \cite{jwc}.   As correlations are short ranged, at not too high density the cluster expansion is expected to be rapidly convergent.

The effective interaction of Ref.~\cite{BV} is 
 derived expanding the left hand side of Eq. \eqref{def:veff}, and keeping the two-body cluster contribution only. The resulting expression is 
\begin{align}
\label{H:2body1}
\langle  H \rangle  = \frac{3 k_F^2}{10 m}  + (\Delta E)_2  \ , 
\end{align}
with
\begin{align}
\label{H:2body2}
(\Delta E)_2 =  \frac{\rho}{2} \int d^3r \   \frac{1}{m} \left[ {\boldsymbol \nabla f(r)} \right]^2  \left[ 1  -  \frac{1}{\nu} \ell^2(k_Fr) \right]  ,
\end{align}
where $\nu$ denotes the degeneracy of the momentum eigenstates, $\rho = \nu k_F^3/6 \pi^2$ is the particle density, and the Slater function is defined as
$\ell(x)~=~3 \left( \sin x - x \cos x \right)/x^3$.
%\end{equation} 

The expression of $v_{\rm eff}$ follows immediately from Eqs.~\eqref{def:veff2}-\eqref{H:2body2}, implying 
\begin{equation}
\label{veff:final}
v_{\rm eff}(r) = \frac{1}{m} \left[ {\boldsymbol \nabla f(r)} \right]^2 \ .
\end{equation}

As pointed out above, the minimisation of $\langle H \rangle$, given by Eqs. \eqref{H:2body1}-\eqref{H:2body2}, yields a Euler-Lagrange equation for  
$f(r)$, to be solved with the boundary conditions $f(a)=0$ and $f(d)~=~1$, and the additional constraint $f^\prime(d)~=~0$, fulfilled introducing a 
Lagrange multiplier.  The role of the correlation range $d$ will be discussed in the next Section.

%%%%%%%%%%%%%%%%%%%%%%%%%%%%%%%%%%%%%%%%%%%%%%%%%%%%%%%%%%%%%%%%%%%%%%%%%%%%%
\subsection{Ground state energy}
\label{energy}
%%%%%%%%%%%%%%%%%%%%%%%%%%%%%%%%%%%%%%%%%%%%%%%%%%%%%%%%%%%%%%%%%%%%%%%%%%%%%

The terms of the cluster expansion can be conveniently represented by diagrams, that can be classified according 
to their topological structure. Selected classes of diagrams can then be summed up to all orders solving a set of
coupled integral equations, dubbed Fermi Hyper-Netted Chain (FHNC) equations \cite{FHNC1,FHNC2}. 

The correlation functions obtained from the procedure outlined above and the FHNC summation scheme have been extensively used to obtain 
upper bounds to the ground state energy of a variety of interacting many-body systems, including liquid helium~\cite{FFM}, nuclear and neutron 
matter~\cite{LBFS} and the Fermi hard-sphere system~\cite{FFPR,AMBP}. Within this approach, yielding remarkably accurate results,  the correlation range 
$d$ is treated as a variational parameter.

\begin{figure}[h!]
%\vspace*{.1in}
 \includegraphics[scale= 0.45]{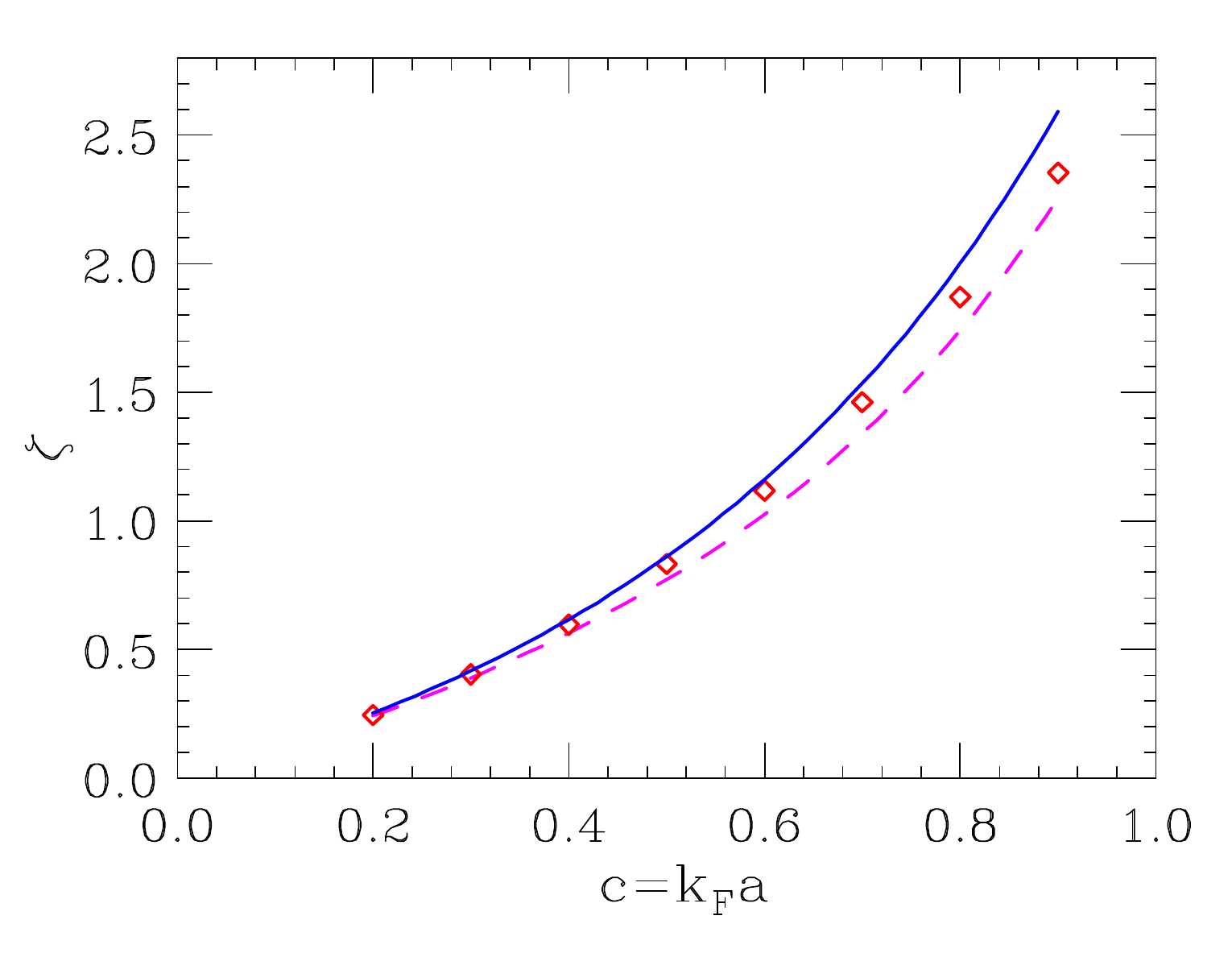}
 \vspace*{-.2in}
\caption{(colour online) The full line shows the $c$-dependence of the dimensionless quantity $\zeta$, defined by Eq. \eqref{def:z},  obtained within the FHNC approach for the system of hard spheres of radius $a=1 \ {\rm fm}$, mass $m = 1 \ {\rm fm}^{-1}$  and degeneracy $\nu=4$. The results obtained from the low-density expansion of 
Eq.~\eqref{pert_en} are represented by the dashed line, 
while the diamonds correspond to the perturbative estimates of $\zeta$ computed neglecting terms of order higher than $c^3$.}
\label{E1}
\end{figure}

Figure \ref{E1} shows the $c$-dependence of the dimensionless quantity $\zeta$, defined by the equation
\begin{equation}
\label{def:z}
E_0 = \frac{3 k_F^2}{10m} \left( 1 + \zeta \right) \ , 
\end{equation}
where $E_0$ denotes the ground state energy of the hard-sphere system. Throughout this article we will consistently set the particle
radius and mass and the degeneracy to the values  $a=1 \ {\rm fm}$, $m = 1 \ {\rm fm}^{-1}$  and $\nu=4$. 

A comparison between the perturbative results obtained from Eq. \eqref{pert_en}  (dashed line) and the FHNC energies (full line) show that at low $c$, corresponding to low density, the 
predictions of the two approaches are very close to one another. At $c=0.2$ (0.3), the difference in $\zeta$ turns out to be less than 5\% (7\%), which translates into an energy difference of less than 1\% (2\%). 
The sizeable discrepancies observed at higher values of $c$ are likely ascribable to the failure of the low-density expansion, although the contributions of cluster terms not 
taken into account within the FHNC scheme may also play a non negligible role.  For reference, we also show, by the diamonds, the perturbative values of $\zeta$ obtained 
including contributions up to order $c^3$. Note that the full line representing the FHNC results is consistently above the perturbative results, and that inclusion of the term of order $c^4 \ln{c}$
leads to a decrease of $\zeta$.  This features suggest that the approximations involved in the FHNC calculation of the ground state expectation value of the hamiltonian do not spoil its upper bound character.

%%%%%%%%%%%%%%%%%%%%%%%%%%%%%%%%%%%%%%%%%%%%%%%
%%%%%%%%%%%%%%%%%%%%%%%%%%%%%%%%%%%%%%%%%%%%%%%%%%%%%%%%%%

The effective interaction is designed to provide an accurate estimate of the ground state
energy at first order of perturbation theory in the Fermi gas basis, which amounts to calculating the expectation value of the hamiltonian in the correlated 
ground state at two-body cluster level. This goal is achieved adjusting the range of the correlation function entering the definition of $v_{\rm eff}$, Eq.~\eqref{veff:final}, 
 in such a way that $\langle H \rangle$, defined by  Eqs.~\eqref{H:2body1}-\eqref{H:2body2}, coincide with the FHNC result.

\begin{figure}[ht]
%\vspace*{.1in}
 \includegraphics[scale= 0.45]{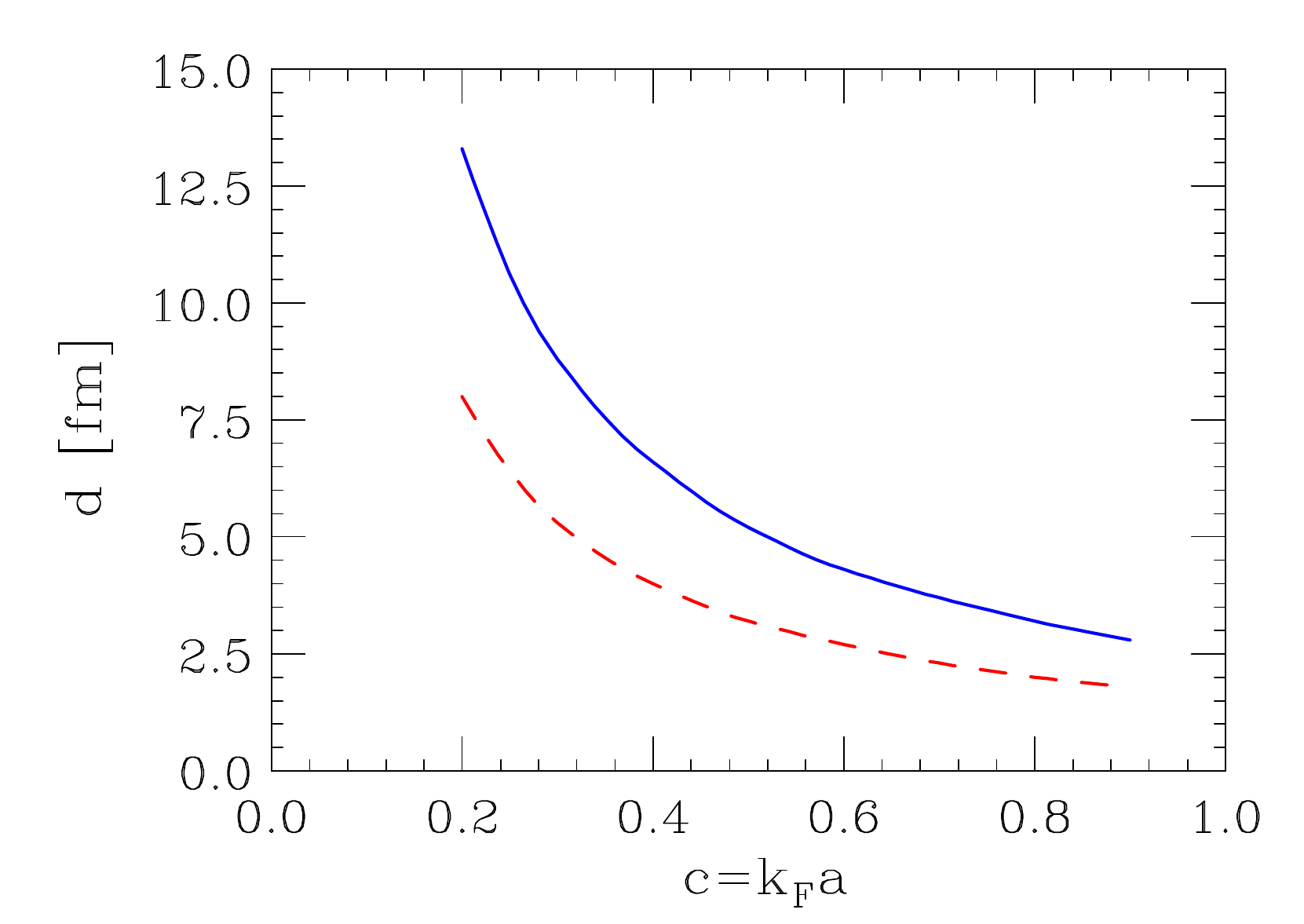}
  \vspace*{-.1in}
\caption{(colour online) The full line shows the $c$-dependence of the correlation range, $d$, resulting from minimisation of the ground state energy of the 
hard-sphere system computed within the FHNC approach.  The dashed line corresponds to the correlation range employed to obtain the CBF 
effective interaction of Eq.~\eqref{veff:final}.}
\label{healing}
\end{figure}

In Fig.~\ref{healing}, the correlation range resulting from minimisation of the FHNC ground state energy is compared to that employed to obtain the CBF effective interaction, as a function of the dimensionless variable $c$. 

%%%%%%%%%%%%%%%%%%%%%%%%%%%%%%%%%%%%%%%%%%%%%%%
%%%%%%%%%%%%%%%%%%%%%%%%%%%%%%%%%%%%%%%%%%%%%%%%%%%%%%%%%%

The range of the effective interaction turns out to be sizeably smaller than the correlation range 
determined by the variational calculation over the whole range of $c$, the difference being $\sim 35 \div 40$\%.
This result is consistent with the observation that the two-body cluster approximation 
consistently underestimates the FHNC energy. Therefore, reproducing the FHNC result at two-body cluster level 
requires a shorter correlation range, leading a stronger effective interaction.

\begin{figure}[ht]
%\vspace*{.1in}
 \includegraphics[scale= 0.40]{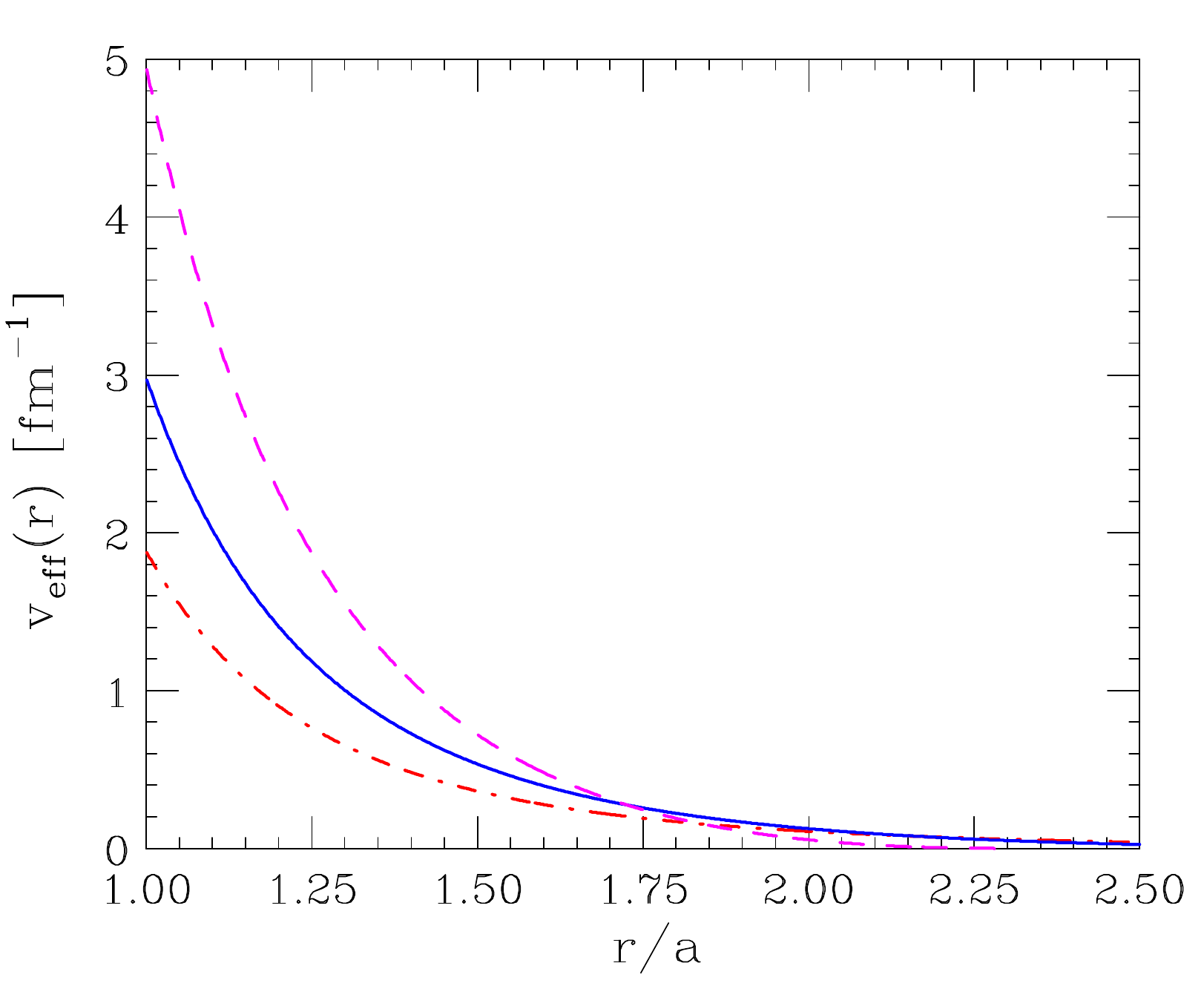}
  \vspace*{-.1in}
\caption{(colour online) Radial dependence of the effective interaction defined by Eq.~\eqref{veff:final}.
The dot-dash, solid and dashed lines correspond to $c=k_Fa=$ 0.3, 0.5 and 0.7, respectively. The region 
$(r/a)<1$, where $v_{\rm eff}(r)=0$, is not shown.}
\label{veff:plot}
\end{figure}

The radial dependence of the effective interaction defined by Eq.~\eqref{veff:final} is illustrated in Fig.~\ref{veff:plot} for 
three different values of the dimensionless variable $c$. Note that the region $(r/a)<1$, where $v_{\rm eff}(r)=0$, is not shown.
The shape of $v_{\rm eff}$ simply reflects the fact that, as the Fermi momentum increases, the correlation range, displayed in Fig.~\ref{healing},  
decreases, and the slope of the correlation function increases. 

%%%%%%%%%%%%%%%%%%%%%%%%%%%%%%%%%%%%%%%%%%%%%%%
%%%%%%%%%%%%%%%%%%%%%%%%%%%%%%%%%%%%%%%%%%%%%%%%%%%%%%%%%%

%%%%%%%%%%%%%%%%%%%%%%%%%%%%%%%%%%%%%%%%%%%%%%%%%%%%%%%%%%%%%%%%%%%%%%%%%%%%%
\section{self-energy}
\label{selfenergy}

The two-point Green's function $G$, embodying all information on single-particle properties of many-body systems, is obtained from Dyson's 
equation \cite{EI1,wimbook}
\begin{align}
\label{dyson}
G(k,E) = G_0(k,E) + G_0(k,E) \Sigma(k,E) G(k,E) \ , 
\end{align}
 where
 % $k = |{\bf k}|$ and 
 $G_0$ is the Green's function of the non interacting Fermi gas, the expression of which reads
\begin{align}
\label{green:FG}
G_0(k,E) = \frac{\theta(k-k_F)}{E - e_0(k) + i \eta} +  \frac{\theta(k_F-k)}{E - e_0(k) - i \eta} \  . 
\end{align}
In the above equation, $\eta = 0^+$,  $e_0(k) = k^2/2m$, $\theta(x)$ is the Heaviside step function, and the two terms in the right hand side 
describe the propagation of particles ($k>k_F$) and holes ($k<k_F$). 
 
The irreducible, or proper, self-energy, $\Sigma(k,E)$, accounts for the effects of interactions. 
From Eqs. \eqref{dyson} and \eqref{green:FG},  it follows  
that in interacting systems the Green's function can be written in terms of the self-energy according to
%the the kinetic energy spectrum, $e_0(k)$, is replaced by the quantity
%\begin{align}
%\epsilon(k,E) = e_0(k)  +  \Sigma(k,E) \ .
%\end{align}
\begin{align}
\label{green:interacting}
G(k,E) = \frac{1}{E - e_0(k) - \Sigma(k,E)} \ .
\end{align}

\begin{figure}[h!]
\vspace*{.1in}
 \includegraphics[scale= 0.65]{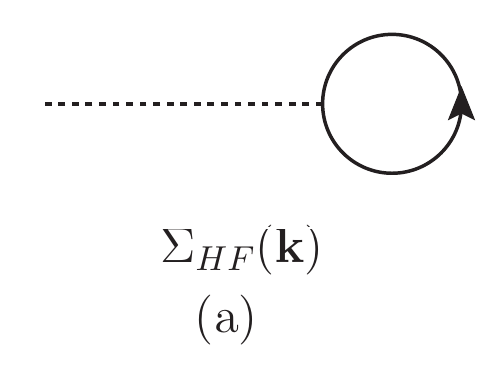}
 \includegraphics[scale= 0.65]{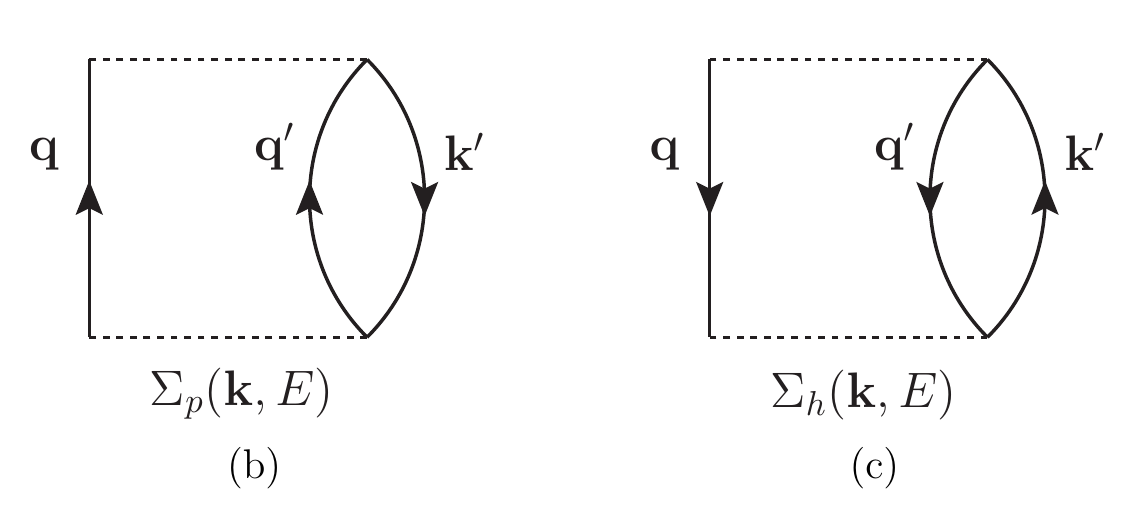}
  \vspace*{-.1in}
\caption{Diagrammatic representation of the direct part of the first and second order contributions to the irreducible self-energy. 
Panels (a), (b) and (c) correspond to the 
Hartree-Fock, polarisation and correlation terms, respectively. Dashed lines represent the CBF effective interaction, while upward and downward oriented 
solid lines depict the  free propagation of  particle and hole states, respectively.}
\label{self}
\end{figure}

In perturbation theory, the irreducible self-energy is obtained from the expansion
\begin{align}
\Sigma(k,E) = \Sigma^{(1)}(k) + \Sigma^{(2)}(k,E) + \ldots   \ , 
\end{align}
where the energy-independent first order term corresponds to the Hartree-Fock approximation, whereas the second order terms,
referred to as polarisation and correlation contributions, describe interaction effects affecting the propagation of particle $(p)$ and hole $(h)$ 
states, respectively. The diagrams representing the direct part of the first and second order contribution to the irreducible self-energy
% $\Sigma^{(1)}(k)$ and $\Sigma^{(2)}(k,E)$
are shown  in Fig. \ref{self}.

%%%%%%%%%%%%%%%%%%%%%%%%%%%%%%%%%%%%%%%%%%%%%%%
%%%%%%%%%%%%%%%%%%%%%%%%%%%%%%%%%%%%%%%%%%%%%%%%%%%%%%%%%%

The real and energy-independent Hartree-Fock contribution is obtained from
\begin{align}
\label{sigma:HF}
\Sigma_{HF}(k) = \frac{1}{\nu}  \sum_{\sigma, {\bf k}^\prime \sigma^\prime} n^0_<(k^\prime) 
\langle {\bf k} \sigma  \ {\bf k}^\prime \sigma^\prime | v_{\rm eff} | {\bf k} \sigma \ {\bf k}^\prime \sigma^\prime \rangle_a  \ ,
\end{align}
where $n^0_<(k) = \theta(k_F - k)$, the two-particle state is antisymmetrised according to $| \alpha \ \beta \rangle_a = (| \alpha \ \beta \rangle - | \beta \  \alpha \rangle )/\sqrt{2}$, 
and the index $\sigma$ labels the discrete quantum numbers specifying the state of a particle carrying momentum ${\bf k}$.
Equations \eqref{sigma_p} and \eqref{sigma_h} show that,  as the effective interaction is diagonal in the space of the discrete quantum numbers, the self-energy does not depend on $\sigma$.

The explicit expression of the polarisation and correlation contributions are (see Fig. \ref{self})
\begin{align}
\label{sigma_p}
\nonumber
\Sigma_p \left( k, E \right)   =  \frac{m}{\nu}   \sum_{\sigma, {\bf k}^\prime \sigma^\prime, \mathbf{q}\tau, \mathbf{q}^\prime \tau^\prime}   & 
\frac{| \langle {\bf q} \tau  \ {\bf q}^\prime \tau^\prime | v_{\rm eff} | {\bf k} \sigma \ {\bf k}^\prime \sigma^\prime \rangle_a|^2}
{q^2 + {q^\prime}^2- {k^\prime}^2  -2m E  -  i \eta}  \\
& \times n^0_{>} (q) n^0_{>}(q^\prime) n^0_{<} (k^\prime) \ , 
\end{align}
and
\begin{align}
\label{sigma_h}
\nonumber
\Sigma_h \left( k, E \right)   =  \frac{m}{\nu}   \sum_{\sigma, {\bf k}^\prime \sigma^\prime, \mathbf{q}\tau, \mathbf{q}^\prime \tau^\prime}   & 
\frac{| \langle {\bf q} \tau  \ {\bf q}^\prime \tau^\prime | v_{\rm eff} | {\bf k} \sigma \ {\bf k}^\prime \sigma^\prime \rangle_a|^2}
{ {k^\prime}^2 - q^2- {q^\prime}^2  + 2m E  -  i \eta}  \\
& \times n^0_{<} (q) n^0_{<}(q^\prime) n^0_{>} (k^\prime) \ , 
\end{align}
with $n^0_>(k) = \theta(k-k_F)$. 
Note that the above definitions imply that ${\rm Im} \Sigma_h \left( k, E>k_F^2/2m \right) = {\rm Im}  \Sigma_p \left( k, E<k_F^2/2m \right)=0$.  

\begin{figure}[h!]
%\vspace*{.1in}
 \includegraphics[scale= 0.45]{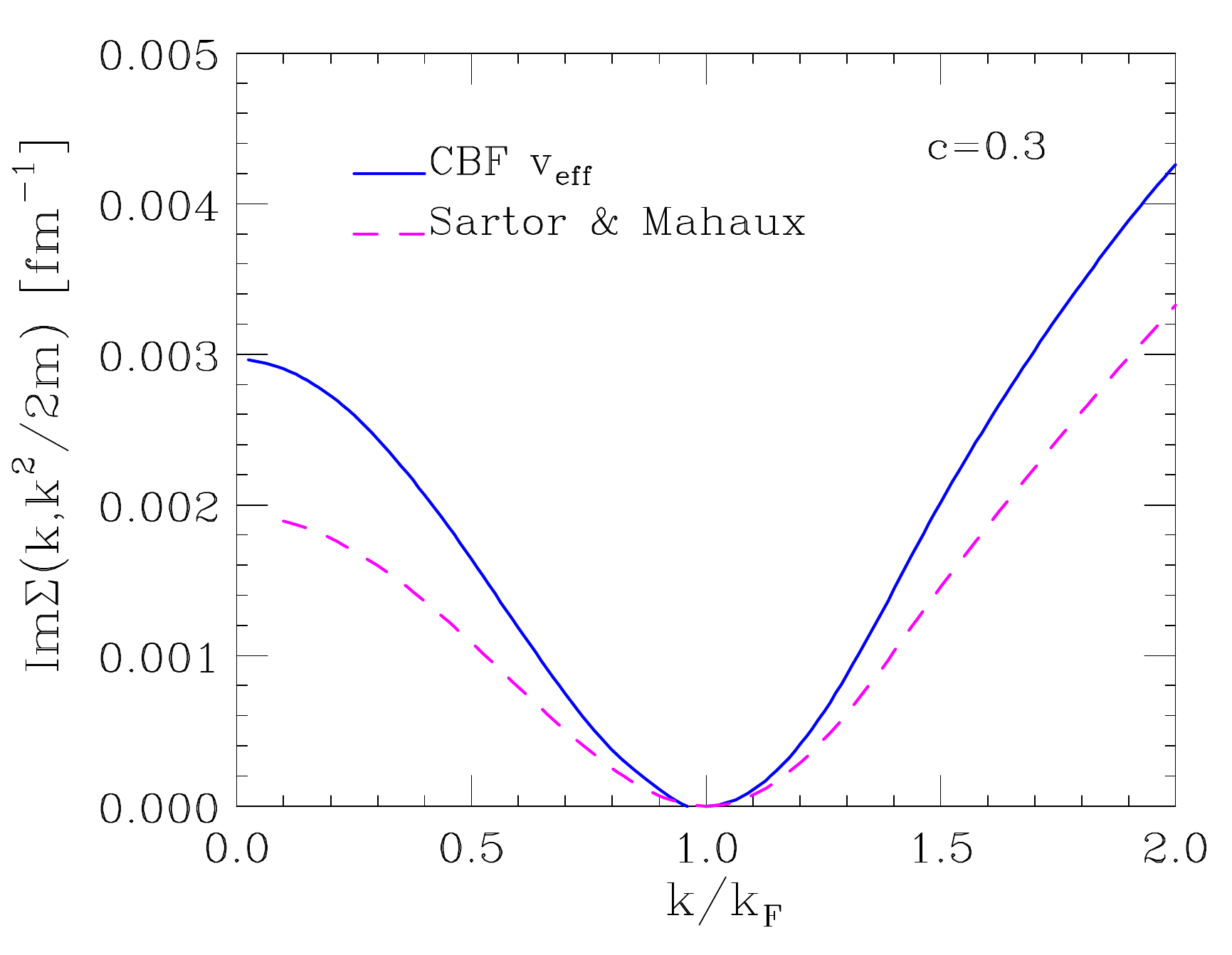}
  \vspace*{-.1in}
\caption{(colour online) Imaginary part of the quantities $\Sigma_h \left( k<k_F, k^2/2m \right)$ and  $\Sigma_p \left( k>k_F, k^2/2m \right)$,
computed at  $c=0.3$ and displayed as a function of the dimensionless variable $k/k_F$. The solid and dashed lines correspond to the 
results obtained from the CBF effective interaction and from the low-density expansion of Ref.~\cite{mahaux}, respectively.}
\label{Im_self}
\end{figure}

Figure \ref{Im_self} shows the behaviour of the imaginary part of $\Sigma_h \left( k, E \right)$ and  $\Sigma_p \left( k, E \right)$ corresponding to 
$c=0.3$, computed at $E = k^2/2m$ and displayed as a function of the dimensionless variable $k/k_F$. For comparison, we also show the same 
quantities computed by Sartor and Mahaux using the low-density expansion and including terms up to order $c^2$ \cite{mahaux}.

%%%%%%%%%%%%%%%%%%%%%%%%%%%%%%%%%%%%%%%%%%%%%%%
%%%%%%%%%%%%%%%%%%%%%%%%%%%%%%%%%%%%%%%%%%%%%%%%%%%%%%%%%%

\begin{figure}[h!]
%\vspace*{.1in}
 \includegraphics[scale= 0.45]{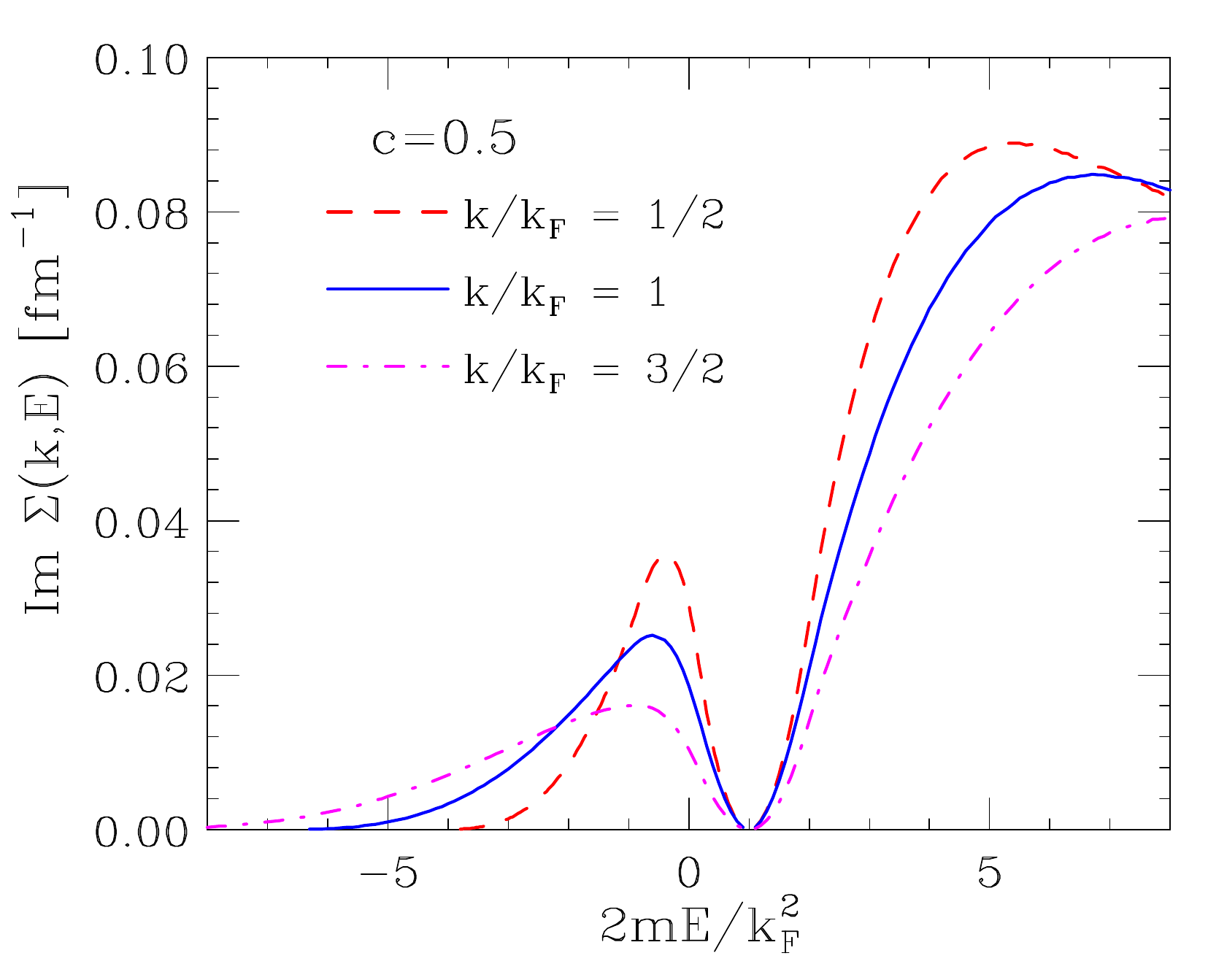}
\vspace*{-.1in}
\caption{(colour online) Energy dependence of the imaginary part of the polarisation ($2mE/k_F^2>1$) and 
correlation ($2mE/k_F^2<1$) contributions to the self-energy of the Fermi hard-sphere system at  $c=0.5$. The dashed, solid 
and dot-dash lines correspond to $k/k_F=$ 0.5, 1 and 1.5, respectively.}
\label{Im_self_off}
\end{figure}

The energy dependence of the imaginary part of the second order contributions to the self-energy is illustrated in Fig. \ref{Im_self_off}, showing the 
results corresponding to $c=0.5$ for three different values of momentum, corresponding to 
$k/k_F=$1/2 (solid line), 1 (dashed line) and 3/2 (dot-dash line).

%%%%%%%%%%%%%%%%%%%%%%%%%%%%%%%%%%%%%%%%%%%%%%%
%%%%%%%%%%%%%%%%%%%%%%%%%%%%%%%%%%%%%%%%%%%%%%%%%%%%%%%%%%

%%%%%%%%%%%%%%%%%%%%%%%%%%%%%%%%%%%%%%%%%%%%%%%%%%%%%%%%%%%%%%%%%%%%%%%%%%%%%
\section{Results}
\label{results}

The self-energy computed at second order in the CBF effective interaction, discussed in the previous Section, has been
used to obtain the single particle spectrum, effective mass and momentum distribution of the 
Fermi hard-sphere system. 

The conceptual framework for the identification of single particle properties in interacting many-body systems
is laid down in Landau's theory of liquid $^3$He (see, e.g. Ref.~\cite{abri}), based on the tenet that there 
is a one-to-one correspondence between the elementary excitations of a Fermi liquid, dubbed quasiparticles, 
and those of the non interacting Fermi gas.

Quasiparticle states  of momentum $k$ are specified by their energy, $e(k)$ and lifetime $\tau_k = \Gamma_k^{-1}$. In the limit of small $\Gamma_k$, the 
Green's function describing  the propagation of quasiparticles  can be written in the form
\begin{align}
\label{green:QP}
G(k,E) = \frac{Z_k}{E - e(k) + i \Gamma_k}  \ . 
\end{align}
A comparison between the above expression and Eq.\eqref{green:interacting} clearly shows that 
quasiparticle properties can be readily related to the real and imaginary parts of the self-energy.

%%%%%%%%%%%%%%%%%%%%%%%%%%%%%%%%%%%%%%%%%%%%%%%%%%%%%%%%%%%%%%%%%%%%%%%%%%%%%
\subsection{Effective mass and single particle spectrum}
\label{mstar}
%%%%%%%%%%%%%%%%%%%%%%%%%%%%%%%%%%%%%%%%%%%%%%%%%%%%%%%%%%%%%%%%%%%%%%%%%%%%%

\begin{figure}[h!]
\vspace*{.1in}
 \includegraphics[scale= 0.55]{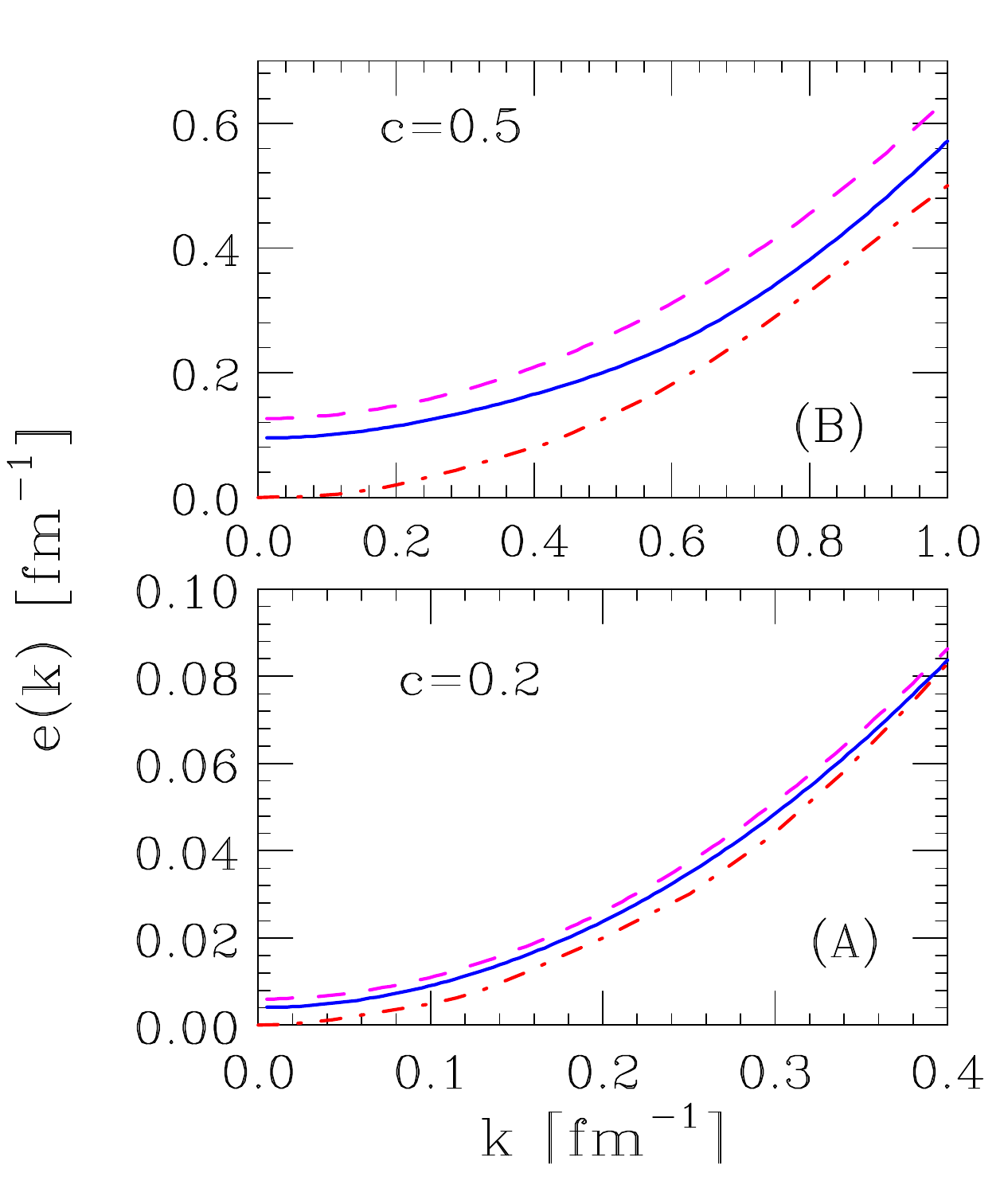}
 \vspace*{-.1in}
\caption{(colour online) quasiparticle energy, computed from Eq. \eqref{def:spectrum} at $c=0.2$ [panel (A)]
and 0.5 [panel(B)]. The dashed and solid lines correspond to the first order (i.e. Hartree-Fock) and second order 
approximations to the self-energy, respectively. For comparison, the dot-dash lines show the kinetic energy spectrum. }
\label{fig:ek}
\end{figure}

The energy of a quasiparticle of momentum $k$, $e(k)$, is obtained solving the equation
\begin{align}
\label{def:spectrum}
e(k) = e_0(k) + {\rm Re} \  \Sigma[k,e(k)] \ .
\end{align}
Substitution of Eq. \eqref{sigma:HF} in Eq.~\eqref{def:spectrum} yields the Hartee-Fock spectrum, represented by 
the dashes lines of Fig. \ref{fig:ek}, while the results obtained including the second order corrections to the self-energy are displayed by
full lines. For comparison, the dot-dash lines show the kinetic energy spectrum.

%%%%%%%%%%%%%%%%%%%%%%%%%%%%%%%%%%%%%%%%%%%%%%%
%%%%%%%%%%%%%%%%%%%%%%%%%%%%%%%%%%%%%%%%%%%%%%%%%%%%%%%%%%

 From Eqs. \eqref{green:interacting} and \eqref{green:QP} it also follows that the quasiparticle lifetime is
related to the self-energy through
\begin{align}
\label{def:lifetime}
%\tau_k^{-1} = \Gamma_k = \left[ \frac{ {\rm Im} \  \Sigma (k,E) }{ 1 + \frac{\partial}{\partial E} {\rm Re} \ \Sigma(k,E)    }  \right]_{E=e(k)} \ .
 \tau_k^{-1} =  \Gamma_k = Z_k {\rm Im} \  \Sigma [k,e(k)] \ , 
\end{align}
where
\begin{align}
\label{residue}
Z_k =  \left[ 1 - \frac{\partial}{\partial E} {\rm Re} \ \Sigma(k,E)  \right]^{-1}_{E=e(k)} \ , 
\end{align}
is the residue of the Green's function of Eq.~\eqref{green:QP} at the quasiparticle pole.

Equations \eqref{def:spectrum} and \eqref{def:lifetime} are obtained expanding the energy of the quasiparticle pole
in powers of $\Gamma_k$, and keeping the linear term only. Note that the resulting expressions of $e(k)$ and $\Gamma_k$
obtained using the second order self-energy are not second order quantities. 

The quasiparticle spectrum is conveniently parametrized in terms of the effective mass $m^\star$, defined by
Eq. \eqref{def:mstar}. 
The total derivative of $e = e(k)$ is performed using Eq.~\eqref{def:spectrum}, and keeping in mind that, 
since ${\rm Re} \  \Sigma(k,E)$ is evaluated at the quasiparticle pole, 
$k$ and $E$ are not independent of one another. As a consequence, one finds
\begin{align}
\frac{d e}{d k} = \frac{k}{m} + \frac{\partial}{\partial k} {\rm Re} \ \Sigma(k,e) 
+ \frac{\partial}{\partial e} {\rm Re} \ \Sigma(k,e) \frac{d e}{d k} \ ,
\end{align}
implying
\begin{align}
\nonumber
 \frac{d e}{d k} & = \left[ \frac{k}{m} + \frac{\partial}{\partial k} {\rm Re} \ \Sigma(k,E) \right]  \\
& \ \ \ \ \ \ \ \  \times  \left[ 1 -  \frac{\partial}{\partial E} {\rm Re} \ \Sigma(k,E)  \right]^{-1}_{E=e(k)} \ .
\end{align}
Note that at first order the self-energy depends on $k$ only, and the above equation reduces to
\begin{align}
\frac{d e}{d k} =   \frac{k}{m} + \frac{\partial  \Sigma_{HF}(k)}{\partial k} \ ,
\end{align}
with $\Sigma_{HF}$ given by Eq. \eqref{sigma:HF}.

\begin{figure}[h!]
\vspace*{.1in}
 \includegraphics[scale= 0.45]{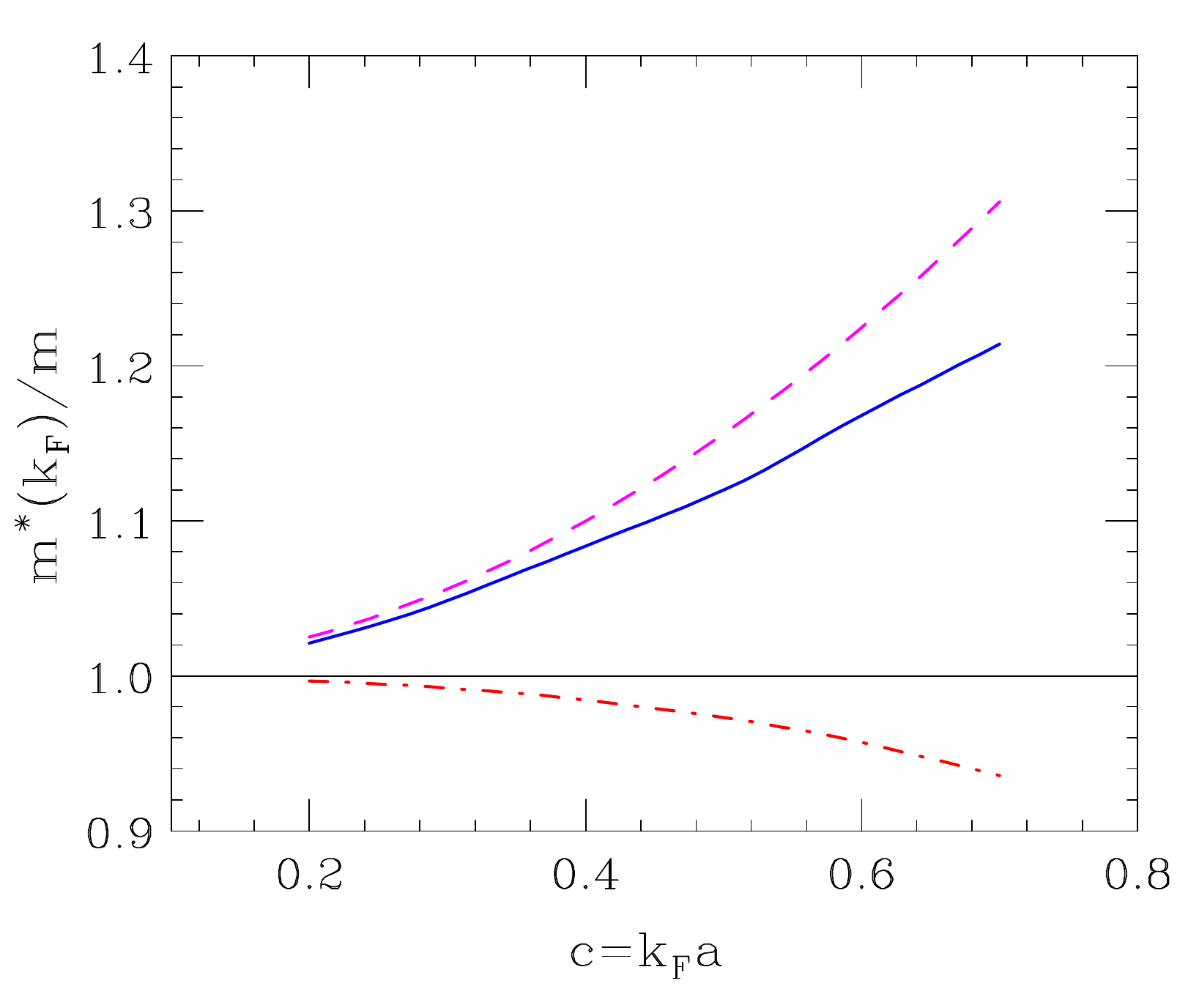}
\caption{(colour online) 
$c$-dependence of the ratio $m^\star/m$ at $k=k_F$, obtained from Eqs.~\eqref{def:spectrum} and \eqref{def:mstar}. 
The dot-dash and solid lines represent the results of 
calculations carried out using the first and second order approximations to the self-energy. 
For comparison, the dashed line shows the results computed using the low-density expansion of Eq. \eqref{pert_mstar}.
}
\label{mstar1}
\end{figure}

The dot-dash and solid lines of Fig.~\ref{mstar1} show the $c$-dependence of the ratio $m^\star(k_F)/m$ at $k=k_F$, evaluated using the self 
energy computed at first and second order in the CBF effective interaction, respectively. It is apparent that inclusion of the energy-dependent contributions to the self-energy, resulting in a moderate correction to the spectra of Fig.~\ref{fig:ek}, 
leads instead to a drastic change in the behaviour of the effective mass. While in the Hartee-Fock approximation the ratio $m^\star(k_F)/m$ is less than one and 
monotonically decreasing with $c$, the full result turns out to be larger than one and monotonically increasing. 

The dashed line of Fig.~\ref{mstar1}, representing the ratio obtained from the low-density expansion at order $c^2$, Eq.~\eqref{pert_mstar}, 
exhibits the same features as the solid line. The low-density expansion, appears to provide quite accurate results for $c \lsim 0.3$. 
A comparison with Fig. \ref{E1} suggests that in the case of the ground state energy the inclusion of higher order contributions extends 
the range of applicability of the expansion to $c \lsim 0.4$.

%%%%%%%%%%%%%%%%%%%%%%%%%%%%%%%%%%%%%%%%%%%%%%%
%%%%%%%%%%%%%%%%%%%%%%%%%%%%%%%%%%%%%%%%%%%%%%%%%%%%%%%%%%

The analysis of the momentum dependence of $e(k)$ reveals that its behaviour is nearly quadratic, and can be accurately
parametrized in terms of $m^\star_F = m^\star(k_F)$ according to 
\begin{align}
e(k) = \frac{k^2}{2m^\star_F} + {\rm const} \ ,
\end{align}
at densities corresponding to $c \lsim 0.5$.
%%%%%%%%%%%%%%%%%%%%%%%%%%%%%%%%%%%%%%%%%%%%%%%%%%%%%%%%%%%%%%%%%%%%%%%%%%%%%
\subsection{Momentum distribution}
\label{momdis}
%%%%%%%%%%%%%%%%%%%%%%%%%%%%%%%%%%%%%%%%%%%%%%%%%%%%%%%%%%%%%%%%%%%%%%%%%%%%%

In translationally invariant systems, the momentum distribution, $n(k)$, describes the occupation probability
of the single-particle state of momentum $k$.  

The connection between $n(k)$ and the Green's function, or the self-energy, can be best understood introducing the 
spectral functions appearing in the Lehmann representation of the two-point
Green's function (see, e.g., Refs.~\cite{wimbook,gottfried})
\begin{align}
\label{KL}
\nonumber
G(k,E)  = \int_0^\infty dE^\prime \ & \left[  \frac{ P_p(k,E)}{E-E^\prime - \mu +i \eta} \right. \\
 &   \ \ \ \ \ \left. + \frac{ P_h(k,E)}{E+E^\prime - \mu - i \eta} \right] \ ,
\end{align}
where $\mu=e(k_F)$ denotes the chemical potential. 

The particle (hole) spectral function $P_p(k,E)$ [$P_h(k,E)$] yields 
the probability of adding to (removing from) the ground state a particle of momentum $k$, leaving the 
resulting $(N+1)$- [$(N-1)$-] particle system with energy $E$. It follows that  
\begin{align}
\label{E:int}
n(k) = \int_0^\infty dE P_h(k,E)  = 1 -  \int_0^\infty dE P_p(k,E) \ .
\end{align}

%%%%%%%%%%%%%%%%%%%%%%%%%%%%%%%%%%%%%%%%%%%%%%%
\begin{figure}[h!]
\vspace*{.1in}
 \includegraphics[scale= 0.45]{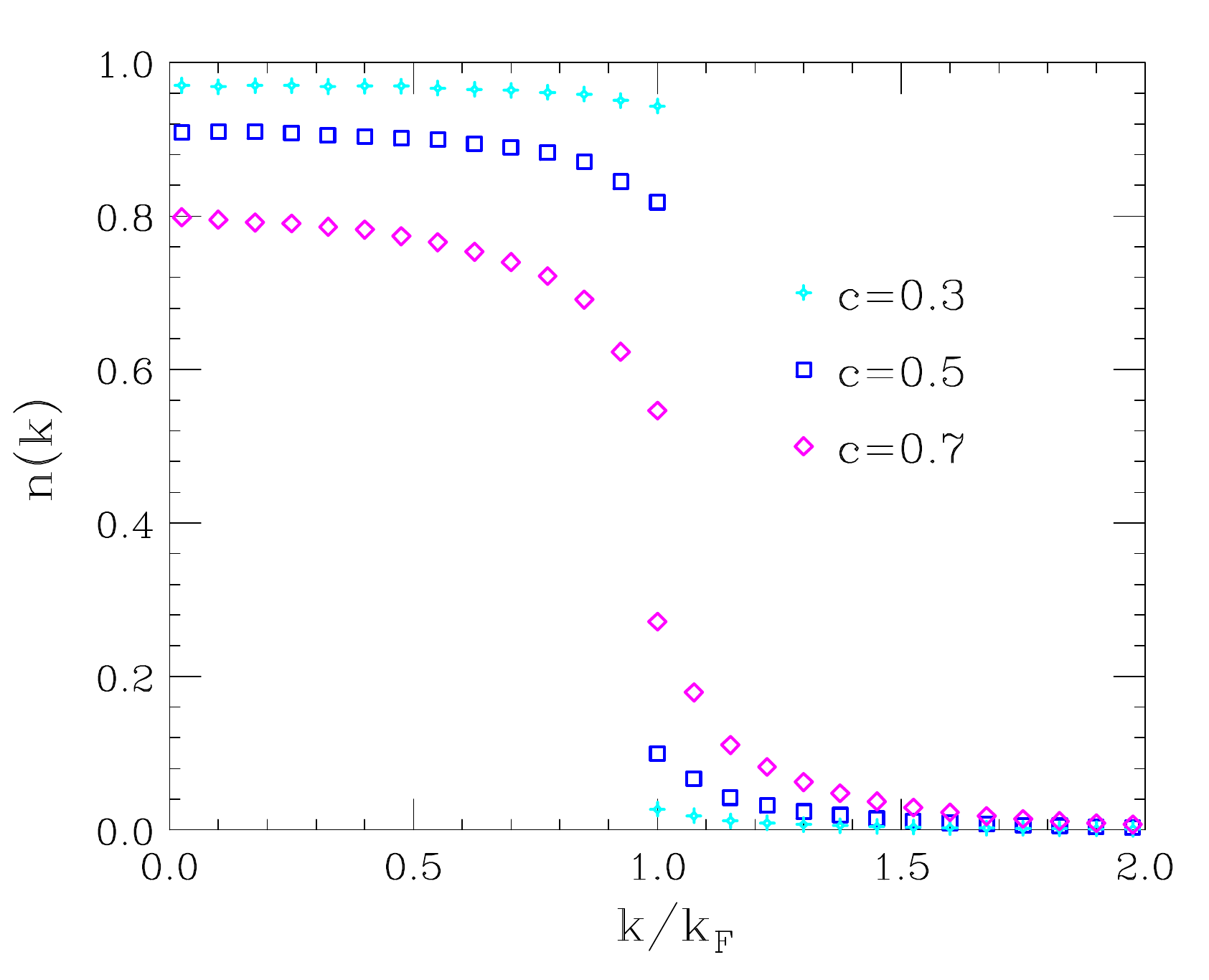}
\caption{(colour online)Momentum distributions computed at second order in the CBF effective interactions, for
three different values of $c=k_F a$. The values of the discontinuity corresponding  $c=$ 0.3, 0.5 and 0.7 are 0.92, 
0.72 and 0.28, respectively.}
\label{nk}
\end{figure}
%%%%%%%%%%%%%%%%%%%%%%%%%%%%%%%%%%%%%%%%%%%%%%%%%%%%%%%%%%

The momentum distribution obtained from Eq. \eqref{E:int}, with 
\begin{widetext} 
\begin{align}
\label{def:Ph}
P_h(k,E) = 
\frac{1}{\pi} {\rm Im} \ G(k,\mu - E)  = \frac{1}{\pi} \frac{ {\rm Im} \Sigma(k,\mu - E) }{ [ \mu - E - e_0(k) - {\rm Re} \Sigma(k,\mu - E) ]^2 + [ {\rm Im} \Sigma(k,\mu - E) ]^2 } \ ,
\end{align}
and
\begin{align}
\label{def:Pp}
P_p(k,E) = 
- \frac{1}{\pi} {\rm Im} \ G(k,\mu + E)  = - \frac{1}{\pi} \frac{ {\rm Im} \Sigma(k,\mu + E) }{ [ \mu + E - e_0(k) - {\rm Re} \Sigma(k,\mu + E) ]^2 + [ {\rm Im} \Sigma(k,\mu + E) ]^2 } \ ,
\end{align}
\end{widetext}
can be cast in the form \cite{gangof3}
\begin{align}
\label{nkpole}
n(k) = Z_k \theta(k_F - k) + \delta n(k) \ .
\end{align}
The first term in the right hand side of the above equation, with $Z_k$ defined by Eq. \eqref{residue}, originates from 
the quasiparticle pole in Eq. \eqref{green:QP}, while $\delta n(k)$ is a smooth contribution,  extending to momenta both below and above $k_F$, 
arising from more complex excitations of the system.  
Equation~\eqref{nkpole} shows that the discontinuity 
of $n(k)$ at $k=k_F$ is given by 
\begin{align}
n(k_F - \eta) - n(k_F + \eta) = Z_{k_F} = Z \ .
\end{align}

At second order in the effective interaction, the momentum distribution obtained from Eqs.\eqref{KL}-\eqref{def:Pp} can be 
conveniently written in the form
\begin{align}
n(k) = n_<(k) + n_>(k) \ ,
\end{align}
where $n_<(k>k_F) = n_>(k<k_F) = 0$, and  
\begin{align}
n_<(k<k_F) & = 1 + \left[ \frac{ \partial}{\partial E} {\rm Re} \Sigma_p(k,E) \right]_{E=e_0(k)} \ , \\
n_>(k>k_F) & = - \left[ \frac{ \partial}{\partial E} {\rm Re} \Sigma_h(k,E) \right]_{E=e_0(k)} \ .
\end{align}
Note that the above equations imply that within the Hartree-Fock approximation $n(k) = \theta(k_F-k)$, and $Z~=~1$.

Figure \ref{nk} shows the momentum distributions obtained including contributions up to second order in the CBF effective interaction, 
for three different values of the dimensionless parameter $c$. It clearly appears that the deviation 
from the Fermi gas result rapidly increases with density. 

\begin{figure}[h!]
\vspace*{.1in}
 \includegraphics[scale= 0.45]{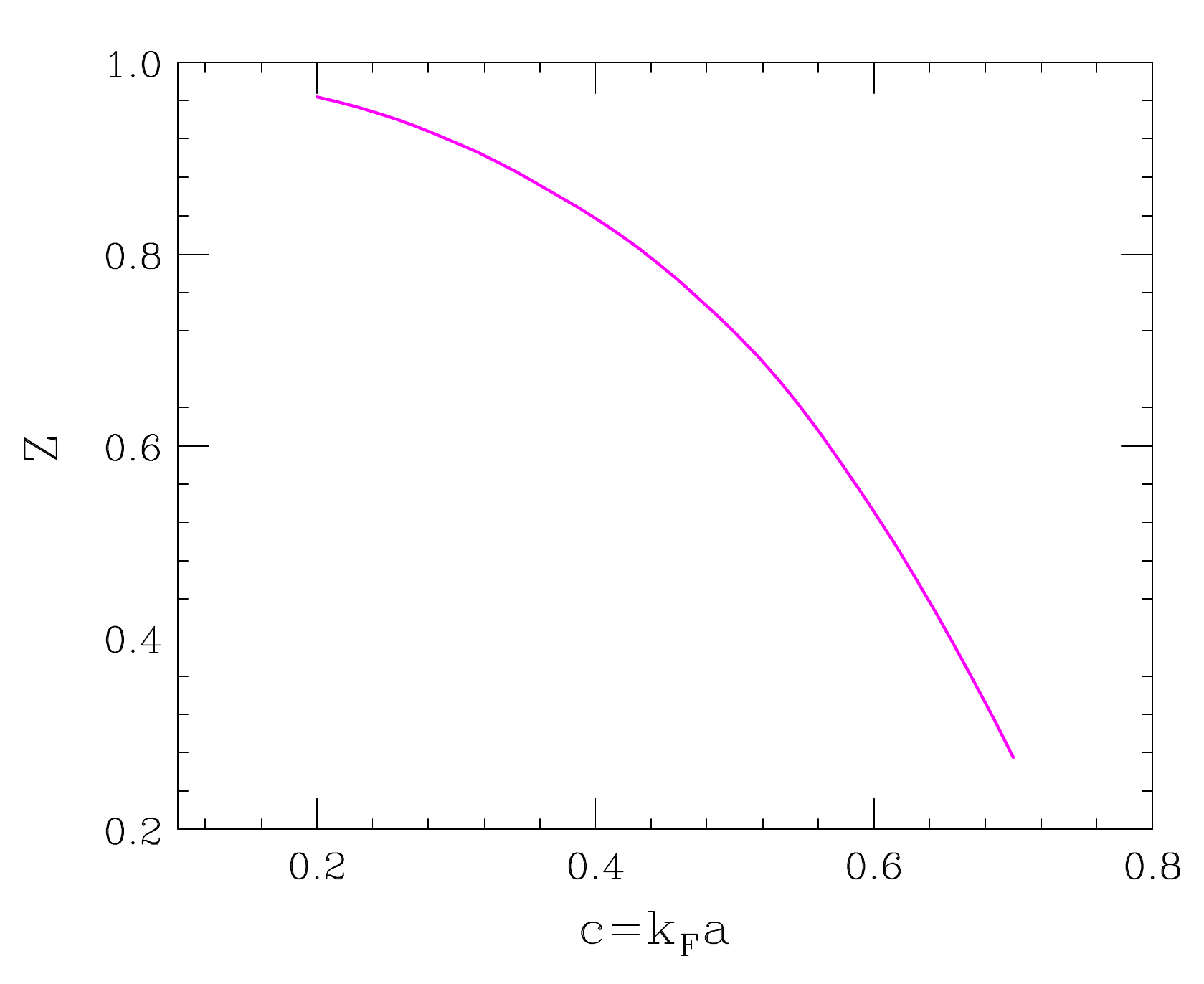}
\caption{(colour online) Discontinuity of the momentum distribution of the Fermi hard-sphere system, as a function 
of $c=k_Fa$.}
\label{z_k}
\end{figure}

A measure of interaction effects is provided by
the discontinuity $Z$, shown in Fig. \ref{z_k} as a function of $c$.

%%%%%%%%%%%%%%%%%%%%%%%%%%%%%%%%%%%%%%%%%%%%%%%
%%%%%%%%%%%%%%%%%%%%%%%%%%%%%%%%%%%%%%%%%%%%%%%%%%%%%%%%%%

\begin{figure}[h!]
\vspace*{.1in}
 \includegraphics[scale= 0.45]{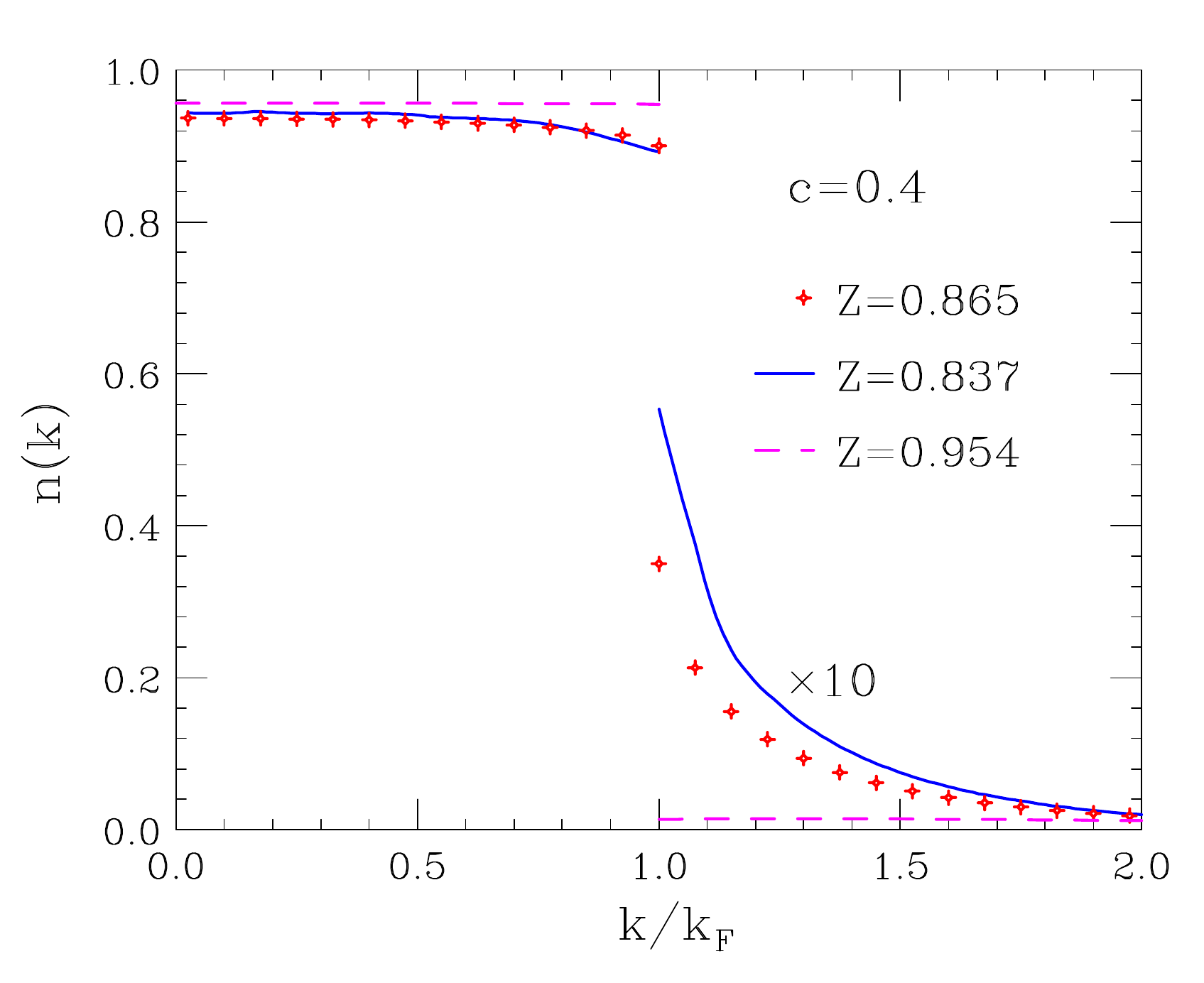}
\caption{(colour online) Momentum distribution of the Fermi hard-sphere system at $c=0.4$. Solid line: results 
obtained at second order in the CBF effective interaction; dashed line: variational results of Ref. \cite{FFPR}; 
crosses: results of the low-density expansion at order $c^2$. All values of $n(k>k_F)$ are multiplied by a factor 10.}
\label{nk_comp1}
\end{figure}

In Fig. \ref{nk_comp1} we compare the momentum distribution resulting from our calculation, represented by the solid line,  
to those reported in Ref. \cite{FFPR} for $c=0.4$. The dashed line shows the results computed using the variational wave function 
obtained from minimisation of the ground state energy within the FHNC scheme, while the crosses correspond to the 
predictions of the  the low-density expansion discussed in 
Refs.\cite{galitskii,mahaux},  including contributions up to order $c^2$. Note that the values of $n(k>k_F)$ are multiplied 
by a factor 10.

%%%%%%%%%%%%%%%%%%%%%%%%%%%%%%%%%%%%%%%%%%%%%%%
%%%%%%%%%%%%%%%%%%%%%%%%%%%%%%%%%%%%%%%%%%%%%%%%%%%%%%%%%%

It clearly appears that the variational approach sizeably underestimates interaction effects, and fails to provide the 
correct logarithmic behaviour at $k$ close to the Fermi momentum. On the other hand, the momentum distributions obtained 
from the CBF effective interaction and from the low-density expansion are in close agreement at $k<k_F$ and exhibit 
discontinuities that turn out to be within $\sim3$\% of one another.

Note that the kinetic energy computed using the variational $n(k)$ exactly agrees with the variational energy.
On the other hand,  the result obtained from the perturbative momentum distribution 
does not necessarily reproduce the kinetic energy calculated using the effective interaction, Eq. \eqref{def:veff}, which  
coincides with the variational energy by definition.  

\begin{figure}[h!]
\vspace*{.1in}
\includegraphics[scale= 0.45]{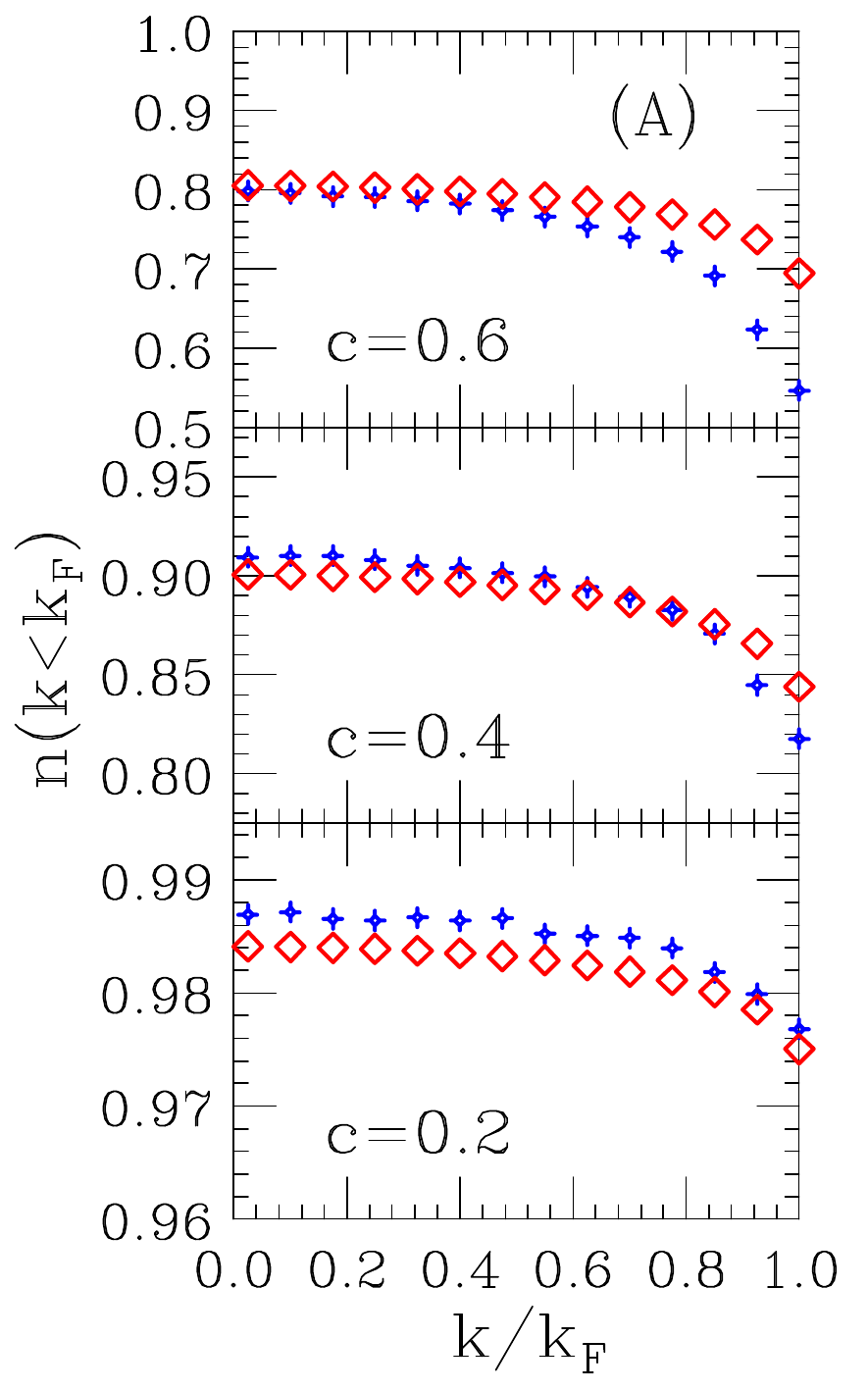}
 \includegraphics[scale= 0.45]{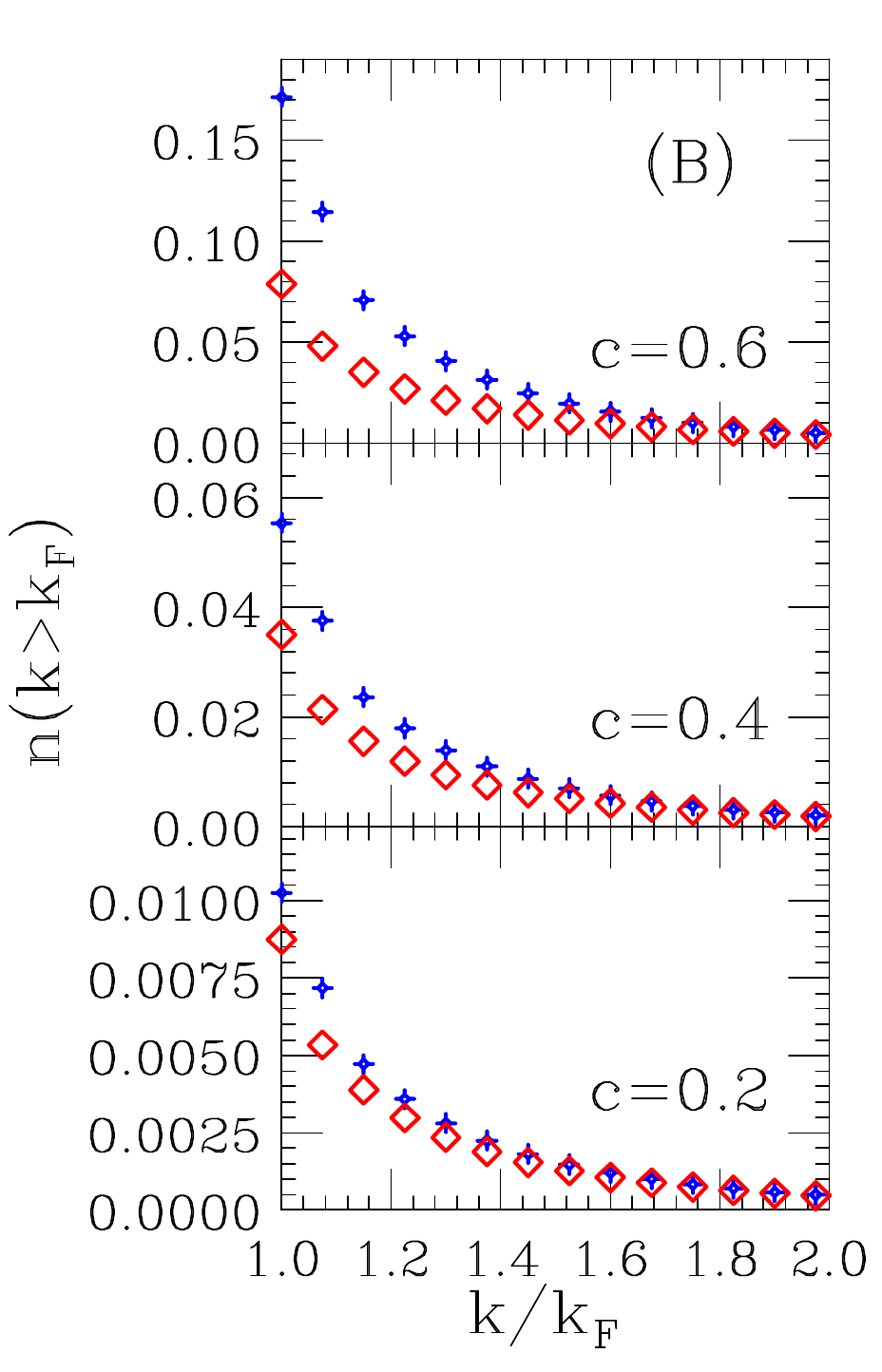}
\caption{(colour online) Comparison between the momentum distribution obtained from the CBF effective
interactions (crosses) and the low-density expansion discussed in Refs.~\cite{galitskii,mahaux} (diamonds), for different
values of the dimensionless parameter $c=k_F a$.  Panels (A) and (B) correspond to the regions $k<k_F$
and $k>k_F$, respesctively.}
\label{nk_comp2}
\end{figure}

In Fig. \ref{nk_comp2}, the difference between the momentum distribution computed using the effective interaction 
and that obtained from the low-density expansion is illustrated for different values of $c$, ranging from 0.2 to 0.6.

%%%%%%%%%%%%%%%%%%%%%%%%%%%%%%%%%%%%%%%%%%%%%%%%%%%%%%%%%%
%%%%%%%%%%%%%%%%%%%%%%%%%%%%%%%%%%%%%%%%%%%%%%%%%%%%%%%%%%

The emerging picture is consistent with that observed in Figs. \ref{E1} and \ref{mstar1}, and suggests that the low density 
expansion provides accurate predictions for $c\lsim0.3$. Sizable discrepancies appear at larger values of $c$, most notably 
in the vicinity of the Fermi surface.

\begin{figure}[h!]
\vspace*{.1in}
 \includegraphics[scale= 0.45]{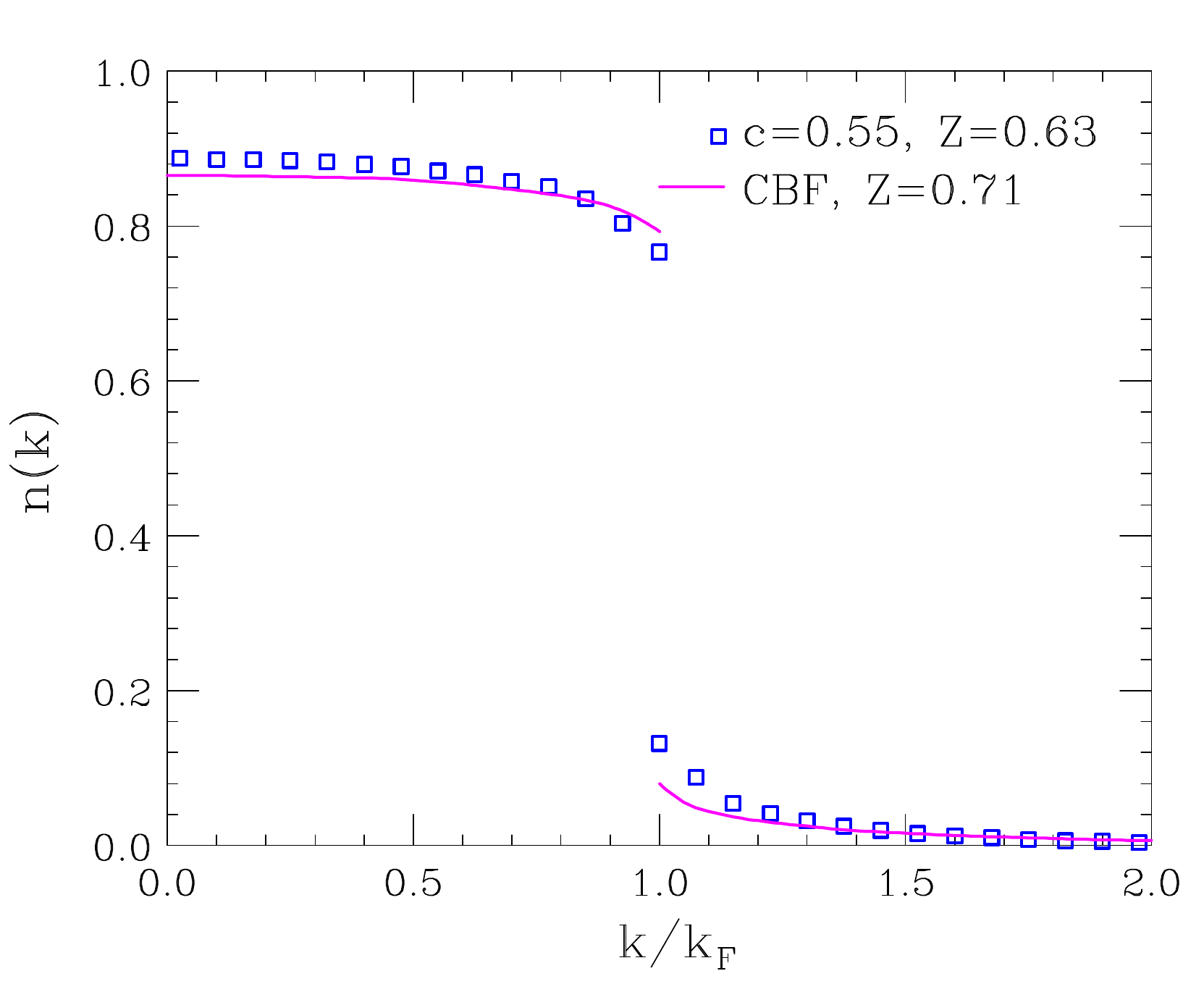}
\caption{(colour online)  Comparison between the momentum distribution obtained from the effective interaction 
approach discussed in this article (squares) and the one  reported in Ref.~\cite{FP} (solid line), computed using correlated a wave functions and second order 
CBF perturbation theory.}
\label{nk_comp_nm}
\end{figure}

In order to establish a correspondence between the hard-sphere system and isospin symmetric nuclear matter at equilibrium density, 
we have analysed the corresponding momentum distributions. In Fig. \ref{nk_comp_nm} the results of our calculations at $c=0.55$ are compared
to the results of the the calculation of Fantoni and Pandharipande \cite{FP}, carried out using a correlated wave function 
and including second order contributions in CBF perturbation theory. Note that the approach of Ref. \cite{FP} is conceptually 
very similar to ours, although the effects of correlations are taken into account modifying the basis states, instead of 
replacing the bare potential with an effective interaction.

%%%%%%%%%%%%%%%%%%%%%%%%%%%%%%%%%%%%%%%%%%%%%%%
%%%%%%%%%%%%%%%%%%%%%%%%%%%%%%%%%%%%%%%%%%%%%%%%%%%%%%%%%%

It appears that, as far as the momentum distribution is concerned,  the system of hard spheres of radius $a=1\ {\rm fm}$ and  
$k_F = 0.55 \ {\rm fm}^{-1}$ corresponds to nuclear matter at density  $\rho_{\rm NM} = 0.16 \ {\rm fm}^{-3}$, or
Fermi momentum $k_F~=~1.33 \ {\rm fm}^{-1}$. Because $n(k)$ is mainly determined by the dimensionless parameter 
$c=k_Fa$,  the results of Fig.~\ref{nk_comp_nm} suggest that nucleons in nuclear matter behave like hard 
spheres of radius $a = 0.55/1.33 \approx 0.4 \ {\rm fm}$. A comparison with nuclear matter momentum distributions 
obtained from other methods \cite{RPD} leads to the same conclusion. It is worth mentioning that the discussion of the 
hard-core model of nuclear matter of Ref. \cite{EI1}, based on the solution of the Bethe-Goldstone equation, also assumes a
hard-core radius $a=0.4 \ {\rm fm}$. 

%%%%%%%%%%%%%%%%%%%%%%%%%%%%%%%%%%%%%%%%%%%%%%%%%%%%%%%%%%%%%%%%%%%%%%%%%
\section{Conclusions}
\label{concl}
%%%%%%%%%%%%%%%%%%%%%%%%%%%%%%%%%%%%%%%%%%%%%%%%%%%%%%%%%%%%%%%%%%%%%%%%%%%%%

We have carried out a perturbative calculation of the properties of the Fermi hard-sphere system using an effective 
interaction derived within the CBF formalism and the cluster expansion technique. 

The proposed approach combines the effectiveness of including correlation effects through a modification of the basis states with the 
flexibility of perturbation theory in the Fermi gas basis. This feature is manifest in the calculated momentum distributions, which, 
unlike those obtained using correlated wave functions in the context of the variational method, exhibit the correct logarithmic shape
in the vicinity of the Fermi surface.  Achieving the same result using the bare interaction and a correlated basis involves 
non-trivial  difficulties, associated with the use of non-orthogonal perturbation theory \cite{FP}. 

The single particle properties obtained from the self-energy computed using the CBF effective interaction turn out to be significantly affected 
by the energy-dependent second order contributions to the self-energy. In the case of the effective mass at momentum $k=k_F$, including these 
contributions leads to a dramatic change of both the values and the density-dependence of the ratio $m^\star(k_F)/m$, with respect to the 
predictions of the Hartree-Fock approximation. Similar results have been found in nuclear matter calculations, carried out within  
G-matrix \cite{Gmatrix,Gmatrix2}, Self Consistent Green's Function \cite{SCGF}  and CBF  \cite{FFP} perturbation theory.

The enhancement of the effective mass has important implications for the calculation of the in medium scattering cross section, which in turn determines the transport coefficients,  
as the value of the effective mass affects both the flux and the phase-space available to the particles in the final state.   

A comparison between the results discussed in this article and those obtained from low-density expansions suggests that the 
latter  provide accurate predictions in the density range corresponding to $k_F \lsim 0.3 - 0.4\ {\rm fm^{-1}}$.
Note that, according to the argument of Section \ref{momdis}, these values of $k_F$ correspond to densities 
in the range $0.2  \lsim (\rho /  \rho_{NM}) \lsim 0.4$, $\rho_{NM}$ being the equilibrium density of 
isospin symmetric nuclear matter.
The analysis of the ground state energy, in which the expansion in powers of $c$ can be tested against accurate 
upper bounds obtained from the variational FHNC approach, suggests that the discrepancies observed a larger $k_F$ are 
likely to be ascribed to contributions of higher order terms in the low-density expansion.

%the accuracy of the calculation of the self-energy at second order in the CBF effective
%interaction needs to be investigated.  

Further insight on the accuracy of the effective interaction approach may be gained 
from its extension to the study of quasiparticle scattering, which has been also investigated 
using Landau's kinetic theory \cite{JLTP1}, as well as of transport properties \cite{JLTP2}. 

In view of applications to dense matter of astrophysical interest, the formalism developed in this article can be readily generalised, 
along the line discussed in Ref. \cite{response}, to obtain a number of properties of isospin-symmetric nuclear matter at equilibrium 
density, such as the spectral functions defined by  Eq. \eqref{KL} and the density and spin-density responses \cite{response}.  
Comparison between the results obtained from the CBF effective 
interaction and those derived from different many-body techniques and using different nuclear 
hamiltonians \cite{GMSF,CBFSF,CBFGREEN,response,baldo1,carbone1} will allow to firmly assess
the potential of this new approach.

%%%%%%%%%%%%%%%%%%%%%%%%%%%%%%%%%%%%%%%%%%%%%%%%%%%%%%%%%%%%%%%%%%%%%%%%%%%%%%
\acknowledgments
%%%%%%%%%%%%%%%%%%%%%%%%%%%%%%%%%%%%%%%%%%%%%%%%%%%%%%%%%%%%%%%%%%%%%%%%%%%%%%
This research is supported by the U.S. Department of Energy, Office of Nuclear Physics, under 
contract DE-AC02-06CH11357 (AL),  MICINN (Spain), under grant FI-2011-24154 and Generalitat de Catalunya, under grant 2014SGR-401 (AP), and  
INFN (Italy) under grant MANYBODY (AM and OB).  AM gratefully acknowledges the hospitality of the 
Departament d'Estructura i Constituents de la Mat\`eria af the University of Barcelona, and support from ÒNewCompStarÓ, 
COST Action MP1304.
%%%%%%%%%%%%%%%%%%%%%%%%%%%%%%%%%%%%%%%%%%%%%%%%%%%%%%%%%%%%%%%%%%%%%%%%%%%%%%

%%%%%%%%%%%%%%%%%%%%%%%%%%%%%%%%%%%%%%%%%%%%%%%%%%%%%%%%%%%%%%%%%%%%%%%%%%%

\end{document}